\documentclass[12pt]{article}

\usepackage{amsmath}
\usepackage{graphicx,psfrag}
\usepackage{enumerate}
\usepackage{natbib}
\usepackage{url} 

\newcommand{\blind}{1}

\begin{document}

\def\spacingset#1{\renewcommand{\baselinestretch}%
{#1}\small\normalsize} \spacingset{1}


\if1\blind
{
  \title{\bf Parameter Restrictions for the Sake of Identification:
    Is there Utility in Asserting that Perhaps a Restriction Holds?}
  \author{Paul Gustafson \thanks{
    The author gratefully acknowledges research support from the Natural Sciences and Engineering Research Council of Canada}\hspace{.2cm}\\
    Department of Statistics\\University of British Columbia}
  \maketitle
} \fi

\if0\blind
{
  \bigskip
  \bigskip
  \bigskip
  \begin{center}
    {\LARGE\bf Parameter Restrictions for the Sake of Identification:
    Is there Utility in Asserting that Perhaps a Restriction Holds?}
\end{center}
  \medskip
} \fi

\bigskip
\begin{abstract}
Statistical modeling can involve a tension between assumptions and statistical identification.
The law of the observable data may not uniquely determine
the value of a target parameter without invoking a key assumption,
and,
while plausible,
this assumption may not be obviously true in the scientific context at hand.
Moreover,
there are many instances of key assumptions which are untestable,
hence we cannot rely on the data to resolve the question of whether the target is legitimately identified.
Working in the Bayesian paradigm,
we consider the grey zone of situations where a key assumption,
in the form of a parameter space restriction,
is scientifically reasonable but not incontrovertible for the problem being tackled.
Specifically,
we investigate statistical properties that ensue if we structure a prior distribution to assert that {\em maybe} or {\em perhaps} the assumption holds.
Technically this simply devolves to using a mixture prior distribution putting just some prior weight on the assumption, or one of several assumptions, holding.
However,
while the construct is straightforward,
there is very little literature discussing situations where Bayesian model averaging is employed across a mix of fully identified and partially identified models.
\end{abstract}

\noindent%
\vfill

\newpage
\spacingset{1.45} 

\section{Introduction}

\subsection{Background}

A tension can arise in applied statistical modeling.
To fully identify a parameter of interest,
``strong'' model assumptions may be required.
Of course it is not prudent to invoke an assumption without a solid rationale.
Without identification,
however,
an uncomfortable truth ensues:
even an infinite amount of data would not reveal the true value of the target parameter.
(Though the data may help some. We shall see
situations of {\em partial} or {\em set} identification,
where an infinite amount of data could rule out {\em some} {\em a priori} plausible values of the target.)
To further muddy the waters,
often an identifying assumption is not testable empirically.
So we cannot necessarily rely on the data to resolve the situation.

Consider a context where an identifying assumption or parameter {\em restriction} is scientifically plausible but not incontrovertible.
One possibility,
at least within the Bayesian paradigm,
is to specify a prior distribution asserting that ``large'' violations of the restriction are unlikely.
For instance, say the restriction is $\lambda=0$.
A prior $\lambda \sim N(0,\tau^{2})$,
for a suitably small choice of $\tau$,
could encode such information.
Some work in this spirit includes
\citet{scharfstein2003incorporating, gugr2006biom, gu2007jrssb, keele2017bayesian, franks2019flexible}.
``Bayesian-like'' approaches, often referred to as {\em probabilistic bias analysis}, have also been considered
\citep[see, for instance][]{greenland2003jasa, greenland2005jrssa, lash2009, lash2014good}.

Assigning a prior distribution which probabilistically limits the magnitude of violation for an identifying restriction is not the only way to proceed.
Arguably a more direct encoding of ``plausible but not incontrovertible'' results from a mixture prior distribution,
giving some weight to the restriction being met exactly, and the remaining weight to it being violated (without necessarily making a stringent judgement about the magnitude of violation).
While mixture prior distributions are ubiquitous in Bayesian hypothesis testing and model selection procedures,
there is scant literature on their use in the context of identifying restrictions.

\subsection{Motivating Example: Prevalence Estimation with Missing Data}

Consider the HIV surveillance study described by
\citet{verstraeten1998pooling},
which also motivated the methodological developments in
\cite{vansteel2006statsinica}.
Blood draws were taken from a sample of $n=787$ members of the target population.
Let
$Y$ indicate the HIV test result ($0$ for negative, $1$ for positive), let $R$ indicate the
observation of the result ($1$ for observed, $0$ for missing),
and let $p_{ry}=Pr(R=r,Y=y)$.
Hence the target parameter,
HIV prevalence,
is $\psi=Pr(Y=1)=p_{01}+p_{11}$.
The study data are summarized by
$c_{10}=699$ negative tests ($R=1,Y=0$),
$c_{11}=52$ positive tests ($R=1,Y=1$),
and
$c_{0+}=36$ missing test results ($R=0$).

\begin{figure}
\begin{center}
\includegraphics[width=4in]{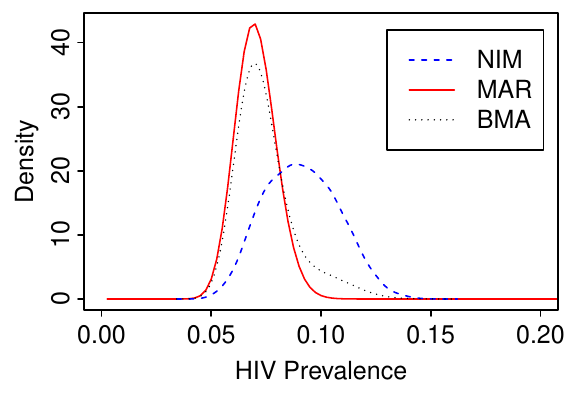}
\end{center}
\caption{Posterior distribution of HIV prevalence without an
identifying restriction (allowing nonignorable missingness, NIM), with the missing-at-random (MAR) restriction, and the Bayesian model averaged
(BMA) synthesis of the two distributions.}
\label{figMAR}
\end{figure}

Consider inferring $\psi$ without invoking any identifying restriction, i.e., allowing there may be nonignorable missingness (NIM).
(See \citet{daniels2008missing} or \citet{little2014statistical} for
full discussions of NIM.)
A weakly informative prior specification is $(p_{00},p_{01},p_{10},p_{11}) \sim \mbox{Dirichlet}(1,1,1,1)$,
i.e., a uniform distribution.
By reparameterizing to $(s,p_{0+},p_{10},p_{11})$,
where $p_{0+}=p_{00}+p_{01}$ and $s=p_{01}/(p_{00}+p_{01})$,
the posterior distribution is characterized by
$(p_{0+},p_{10},p_{11}) \sim \mbox{Dirichlet}(2+c_{0+},1+c_{10},1+c_{11})$
and
independently
$s \sim \mbox{Unif}(0,1)$.
This induces the marginal posterior distribution
on the target $\psi=sp_{0+} + p_{11}$.
The impact of not having an identifying restriction is clear.
Since {\em a posteriori} $s \sim \mbox{Unif}(0,1)$ for any dataset,
the posterior distribution of $\psi$ will not reduce to a point-mass in  the infinite limit of further data collection.
For the data at hand,
the posterior distribution of $\psi$ is depicted in Figure \ref{figMAR}.

Alternately, say we believe the missing-at-random (MAR) assumption is justified, i.e., presuming $R$ to be independent of $Y$.
A weakly informative prior specification is
to let $p_{ry}=(1-\gamma)^{1-r}\gamma^{r}(1-\psi)^{1-y}\psi^{y}$, with
$\gamma=Pr(R=1)$ and $\psi=Pr(Y=1)$ independently and identically distributed as $\mbox{Unif}(0,1)$.
This yields the
marginal posterior distribution of the target $\psi$ as simply
$\mbox{beta}(1+c_{11},1+c_{10})$.
Clearly this posterior would concentrate to a point-mass in the infinite limit of further data collection.
For the data at hand,
this posterior is also depicted in Figure \ref{figMAR}.

An investigator believing that the MAR assumption {\em might} hold could
apply {\em Bayesian model averaging} (BMA)
\citep[see, for instance,][]{kass1995bayes, hoeting1999bayesian, wasserman2000bayesian}.
Applied here,
the prior is simply taken as a mixture of the two specifications above.
The conjugate prior specifications give the Bayes factor contrasting the two specifications (with MAR in the numerator
versus NIM in the denominator) as:
\begin{eqnarray} \label{bf}
b &=& \frac{(n+2)(n+3)}{6(c_{0+}+1)(n-c_{0+}+1)},
\end{eqnarray}
which evaluates to $b=3.73$ for the present data.
And the posterior odds favoring MAR are $b$ times the prior odds.
For instance,
if the mixture prior gives equal weights of $0.5$ to each of the NIM and MAR specifications,
then
for the present data the mixture posterior gives weight $0.211$ to the NIM posterior and weight $0.789$ to the MAR posterior.
This model-averaged posterior for the target is also depicted in Figure \ref{figMAR}.

With the NIM and MAR prior specifications being only weakly informative,
a Bayes factor of nearly four might seem surprising,
as MAR is well known to be an untestable assumption.
Similarly,
the strong dependence of (\ref{bf}) on the proportion of data missing is curious.
However,
oftentimes Bayes factors are sensitive to within-model prior specifications.
In Appendix A we elaborate on this point.

The findings above beg more general questions.
While there is a rich literature on BMA,
averaging across models which do and don't identify the target parameter has not been addressed.
Thus we seek to understand statistical performance
when model averaging is used to declare the {\em a priori} supposition that {\em perhaps} an identifying restriction holds.

\section{Theory and Methods}

\subsection{Framework}

Let $M_0$ be the ``unrestricted'' statistical model at hand,
which lacks full identification.
The initial declaration of $M_0$ is presumed to use parameters which are as scientifically meaningful as possible.
In Bayesian terms it is likely natural to specify a prior distribution in terms of this initial ``scientific'' parameterization.
We presume,
however,
that $M_0$ can then be reparameterized using what \citet{gustafson2015bayesian} refers to as a {\em transparent} parameterization.
This yields parameters $\theta=(\phi,\lambda)$,
such that the distribution of the observable data $D$ depends on $\theta$ only through $\phi$.
Or,
more directly in Bayesian terms,
$D$ is conditionally independent of $\lambda$,
given $\phi$.
The lack of full identification is thus clear:
the data inform $\phi$
(indeed we presume fully, in the sense that
the $(D|\phi)$ model supports consistent estimation of $\phi$).
However,
for any dataset,
the posterior conditional distribution of $(\lambda|\phi)$ is the same as the prior conditional distribution.
To connect with the motivating example of Section 1.2,
$(p_{00},p_{01},p_{10},p_{11})$ is a scientifically interpretable parameterization, but to understand the lack of identification
$\phi=(p_{0+},p_{10},p_{11})$ and $\lambda=s=p_{01}/p_{0+}$ is a transparent parameterization.
Moreover, the $\mbox{Dirichlet}(1,1,1,1)$ prior distribution is declared in the scientific parameterization,
hence the prior distribution on $\theta=(\phi,\lambda)$ follows by change-of-variables.

In what follows,
the primary (scalar) target of inference is expressed as $\psi =g(\phi,\lambda)$.
Provided that $g()$ varies non-trivially with $\lambda$,
the target is not fully identified.
(This is the case in the motivating example,
with $\psi=sp_{0+} + p_{11}$.)

Let $\theta \in \Theta_0$ denote the $M_0$ parameter space,
with the marginal spaces thus implied being $\phi \in \Phi_{0}$ and $\lambda \in \Lambda_{0}$.
Importantly,
many partially identified models arising in practice are such that $\phi$ and $\lambda$ are {\em not} variation independent,
i.e., $\Theta_0$ is {\em not} the Cartesian product space of $\Phi_0$ and $\Lambda_0$.
Or,
more plainly,
the possible values of $\lambda$ depend on $\phi$,
and vice-versa.
This can arise in the move from a scientific parameterization (likely with a Cartesian parameter space) to a transparent parameterization.
In such situations,
direct learning about $\phi$ from the observed data may induce some indirect learning about $\lambda$,
via the support of $\lambda$ depending on $\phi$.
In non-Bayesian settings,
this manifests as bounds on $\lambda$   (and in turn bounds on the target parameter) that are functions of $\phi$,
and therefore are fully estimable \citep[see, for instance,][]{manski2003}.
In the Bayesian setting, this manifests as necessarily having prior dependence between $\phi$ and $\lambda$,
so that the posterior marginal distribution of $\lambda$ can depend on the data, even when the posterior conditional distribution of $(\lambda|\phi)$ does not.
This situation is described at length in \citet{gustafson2015bayesian}.

We consider one or more (in general $J$) sub-models or restrictions
of $M_{0}$,
each of which (i), has some level of {\em a priori} scientific credence,
and (ii), identifies the target.
We label the $j$-th such sub-model as $M_{j}$, for $j=1, \ldots J$, and can regard the
restriction as $\theta \in \Theta_{j} \subset \Theta_{0}$.
Since the sub-model identifies the target,
over $\Theta_{j}$, $g(\phi,\lambda)$ does not vary with $\lambda$.
More pragmatically,
if $M_{j}$ holds,
then $\phi$,
which is estimable,
uniquely determines the target $\psi$.

As already mentioned    ,
the BMA framework lets the user postulate that {\em perhaps} one of the identifying restrictions holds.
Let $\pi_{j}()$ be the $j$-th prior density for $\theta$ over $\Theta_0$.
This will generally be a ``degenerate'' density for $j\geq 1$, since each identifying restriction forces $\theta$ onto a lower-dimensional subset of $\Theta_0$.
Fortunately, the BMA framework is particularly well-suited to fairly compromising between
a larger model and a nested sub-model \citep[see, for instance,][]{kass1995bayes}.
Indeed,
the conceptual symmetry with which a smaller and larger model are treated in a Bayes factor is generally a point of appeal.

The BMA prior distribution is a mixture, which we can represent as:
\begin{eqnarray} \label{BMA prior}
\pi_{MIX}(\theta) &=& \sum_{j=0}^{J} w_{j} \pi_{j}(\theta),
\end{eqnarray}
where $w_{j} = Pr(M_{j})$ is the prior probability that model $M_{j}$ is correct,
hence $\sum_{j=0}^{J} w_{j} =1$.
To be clear, $(w_{1},\ldots,w_{J})$ are hyperparameters which must be specified by the user.
For instance,
if $J=1$, then a setting of $w=(0.5, 0.5)$ gives equal credence to the identifying restriction holding or not holding.

Upon receipt of data $D$,
standard Bayesian updating from the mixture prior (\ref{BMA prior}) gives
the posterior distribution as a mixture:
\begin{eqnarray*}
\pi_{MIX}(\theta | \mbox{D}) &=&
\sum_{j=0}^{J} \tilde{w}_{j}(D) \pi_{j}(\theta|D).
\end{eqnarray*}
Here $\pi_{j}(\theta| D)$ is the standard ``within-model'' posterior distribution of parameters,
i.e., based on
$\pi_{j}(\theta|D) \propto f(D | \theta)\pi_{j}(\theta)$.
And $\tilde{w}_{j}(D)$ is the posterior probability that $M_{j}$ is correct.
In the usual fashion,
\begin{eqnarray*}
\tilde{w}_{j}(D) &=&   \frac{w_{j} f_{j}(D)} {\sum_{k=0}^{J}w_{k} f_{k}(D)},
\end{eqnarray*}
where $f_{j}(D) = \int_{\Theta_{0}} f(D | \theta) \pi_{j}(\theta) d\theta$ is the marginal density of the data under the $j$-th model.

While the framework above is standard,
its implications are unexplored when the list of candidate models is a mix of partially and fully identified models.
To large extent,
the ensuing statistical behavior will be seen to be specific to the
particular choice of partially identified model and identified sub-models,
and to the nature of the specified within-model priors.
However,
to foreshadow the examples which follow,
there are situations where:
\begin{itemize}
\item
{\em Scenario 1.}
For every dataset,
$\tilde{w}_{1}(D)/\tilde{w}_{0}(D) =w_{1}/w_{0}$,
i.e., the data have zero ability to discriminate between the base model and an identified sub-model.

\item
{\em Scenario 2.}
Let $D_n$ denote a dataset of $n$ observations
arising under a particular parameter value $\theta=\theta^{\dagger}$,
and let $w_{1}^{\star}(\theta^\dagger)$ be the large-sample limit of $\tilde{w}_{1}(D_n)$.
It can happen that the restriction is fully testable, in that $w_{1}^{\star}(\theta^\dagger) = I_{\Theta_1}(\theta^{\dagger})$,
i.e., in the large-sample limit one correctly proves or refutes that the restriction holds.

\item
{\em Scenario 3.}
It can happen that the range of $w_{1}^{\star}()$ over $\Theta_0$ is an interval
$[c,1)$,
for some $c \in (0, w_{1})$.
Since
the limiting posterior weight on $M_1$ is never zero or one,
the restriction can never be fully refuted or fully supported.   Also, since $c$ is positive,
there are not parameter values under which we get arbitrarily close to
complete refutation
of the restriction.
In the other direction,
however,
there are parameter values under which we get arbitrarily close to full support {\em for} the restriction.

\item
{\em Scenario 4.}
For data generated under {\em some} $\theta^{\dagger}$ values in $\Theta_{0} - \Theta_{1}$,
$w_1^{*}(\theta^{\dagger})=0$, while for other such values $w_1^{*}(\theta^{\dagger}) \in (0,1)$.
That is, a falsely asserted restriction may or may not be fully refuted.

\end{itemize}
Thus we find a richness in the variety of behaviors that can be encountered.
Indeed, Scenarios 3 and 4 might surprise some,
since they point to nuance beyond a binary construct of untestable versus testable restrictions.

\subsection{Elucidation of Structure}

Towards understanding the posterior model weights,
note that, for $j=0,\ldots , J$,
\begin{eqnarray}
f_{j}(D)  &=  & \int_{\Theta_{0}} f(D | \theta) \pi_{j}(\theta) d\theta  \nonumber \\
&=& \int_{\Theta_{0}} f(D | \phi) \pi_{j}(\phi,\lambda )  d\lambda \; d\phi \nonumber \\
&=&   \int_{\Phi_{0}} f(D | \phi) \pi_{j}(\phi)  d\phi. \label{margform}
\end{eqnarray}
The marginal prior distribution of $\phi$ thus plays a key role.

As was introduced informally in Section 2.1, let $D_{n}$ be $n$ independent datapoints generated
under true parameter values
$(\phi,\lambda) =(\phi^{\dagger}, \lambda^{\dagger})$.
(When helpful,
we use the `dagger' notation to emphasize specific, fixed parameter values giving rise to the observable data.)
Importantly,
note there is a true value of $\lambda$,
even though this doesn't influence the distribution of the observable data.
We presume sufficient regularity for
$(D_{n}|\phi)$ such that standard asymptotic behavior based on Fisher information applies to maximum likelihood estimation of $\phi$ \citep[see, for instance,][]{casella2002statistical}.
Additionally, we presume sufficient smoothness of marginal prior densities
$\pi_{j}(\phi)$ for each $j$, such that the asymptotic normality of the MLE for $\phi$ extends to asymptotic normality of the posterior $\pi_{j}(\phi|D_{n})$ \citep[see, for instance,][]{chen1985asymptotic}.

As introduced in Section 2.1.,
let $w^{\star}_{j} = \lim_{n \rightarrow \infty}  \tilde{w}_{j}(D_{n})$ be the
limiting posterior weight on the $j$-th model,
in the sense of almost sure convergence with respect to the
distribution of $(D_n|\phi^{\dagger})$.
Toward characterizing $w^{\star}$,
from (\ref{margform}) we have, for $j \neq k$,
\begin{eqnarray}
\frac{f_{j}(D_n)}{f_{k}(D_n)}  & =  &
\frac{ \int_{\Phi_{0}} \{\pi_{j}(\phi)/\pi_{k}(\phi)\} f(D_n | \phi) \pi_{k}(\phi) \; d\phi}
{ f_{k}(D_n) } \nonumber \\
& = & E_{k}\{\pi_{j}(\phi)/\pi_{k}(\phi) |D_n\}, \label{margform_ratio}
\end{eqnarray}
where the subscript $k$ reminds us that this posterior expectation is taken with respect to prior $\pi_{k}$.
Now presume that
$(\phi^{\dagger},\lambda^{\dagger}) \in  \Theta_{k}$
(so that (\ref{margform_ratio}) is a posterior expectation arising in a ``right model'' scenario),
and presume that the prior density ratio $\pi_{j}(\phi) / \pi_{k}(\phi)$ is finite and continuous at $\phi^{\dagger}$.
Then (\ref{margform_ratio})
leads to
\begin{eqnarray}
\frac{w^{\star}_{j}}{w^{\star}_{k}} &=&
\left\{ \lim_{n \rightarrow \infty}  \frac{ f_{j}(D_{n})}{f_{k}(D_{n}) } \right\}
\left(\frac{w_{j}}{w_{k}}\right) \nonumber \\
& = &
\left\{ \frac{ \pi_{j}(\phi^{\dagger})}{\pi_{k}(\phi^{\dagger}) } \right\}
\left(\frac{w_{j}}{w_{k}}\right). \label{genlimbf}
\end{eqnarray}
As a further technical note, should the prior density ratio $\pi_{j}(\phi) / \pi_{k}(\phi)$ not be finite at $\phi^{\dagger}$,
we can interchange $j$ and $k$ before appealing to (\ref{genlimbf}).
Or (\ref{genlimbf}) could be made more formal by replacing the ratio of prior densities with the Radon-Nikodym derivative of one prior distribution with respect to the other.

Just as we can characterize the limiting posterior weights on models,
we can also characterize the limiting values of within-model posterior means.
Generically, let
\begin{eqnarray*}
\hat{\psi}(D ; \pi_{I}) &=&
\int g(\phi,\lambda) \pi_{I}(\phi,\lambda \mid D) \; d\phi \; d\lambda
\end{eqnarray*}
be the posterior mean of the target $\psi$ arising from the prior specification $\pi_{I}$ and data $D$,
where the $I$ subscript reminds us this is the {\em investigator's} choice of prior distribution.
It is easy to verify that without an identifying restriction we have
the large-sample limit of the posterior mean (again in the almost-sure sense
with respect to the law of $(D_n | \phi^{\dagger})$) given as
\begin{eqnarray}
\psi^{\star}_{0} & \doteq &
\lim_{n \rightarrow \infty} \hat{\psi}(D_{n} ; \pi_{0}) \nonumber \\
&=&
\lim_{n \rightarrow \infty}
\int g(\phi, \lambda) \pi_{0}(\lambda | \phi) \pi_{0} (\phi | D_{n}) \; d\lambda \; d\phi  \nonumber \\
\label{limzero}
& = &
\int g(\phi^{\dagger},\lambda) \pi_{0}(\lambda|\phi^{\dagger}) d\lambda.
\end{eqnarray}

We can similarly define limiting posterior means when the investigator imposes
an identifying restriction, i.e., $\psi^{\star}_{j} \doteq \lim_{n \rightarrow \infty} \hat{\psi}(D_{n} ; \pi_{j})$
for
$j>0$.
Specifically,
when the investigator invokes the $j$-th restriction,
inference is based on an identified model parameterized by $\phi$.
If the restriction is valid, the posterior mean of $\psi$ will converge
to the correct value of the target,
i.e., formally $\psi^{\star}_{j}= g_j(\phi^{\dagger})$,
where $g_{j}(\phi)$ is defined as the constant value of $g(\phi,\lambda)$ arising for any $\lambda$ such that $(\phi,\lambda) \in \Theta_{j}$.
If the restriction is not valid, then the limiting posterior mean will converge
to $g_{j}()$ evaluated at the ``pseudo-true'' value of
$\phi$ which minimizes the Kullback-Leibler divergence between the actual distribution of $D$ and the modelled distribution of $(D|\phi)$, as per the wrong-model asymptotic theory of \citet{white1982maximum}.
%

%
%

Assembling the pieces thus far,
the large-sample limit of the BMA posterior mean is
\begin{eqnarray} \label{limmix}
\psi^{\star}_{MIX} & \doteq &
\sum_{j=0}^{J}  w^{\star}_{j} \psi^{\star}_{j}.
\end{eqnarray}

\subsection{Quantifying Performance}

Regarding the posterior mean $\hat{\psi}(D; \pi_{I})$ as an estimator (in this case with a Bayesian motivation) of $\psi$,
we quantify estimation accuracy starting with the usual
frequentist mean-squared error (MSE),
i.e.,
averaging over repeated sampling of $D$ for a fixed $\theta$ to obtain
$MSE(\theta)= E_{\theta}\left[ \{ \hat{\psi}(D; \pi_{I}) - \psi(\theta)\}^{2}\right]$.
(Even though the distribution of $(D|\theta)$ depends on $\theta$ only through $\phi$,
the MSE generally depends on {\em all} of $\theta$,
since the target parameter
$\psi$ will be a function of both $\phi$ and $\lambda$.)
Then we average $MSE(\theta)$ across different values of $\theta$,
to reflect aggregate performance across different scenarios of
what the truth might be.
This averaging is weighted
according to a density $\pi_{N}()$ over $\Theta_0$.
We find it useful heuristically to label $\pi_{N}()$ as {\em Nature's} prior distribution (hence the subscript $N$), as distinct from
$\pi_{I}$ which the investigator uses in performing the Bayesian analysis of the data.
For data $D_{n}$ with sample size $n$,
then,
the {\em average} mean-squared error (AMSE) is defined as:
\begin{eqnarray} \label{AMSE}
AMSE_{n}(\pi_{N},\pi_{I})
& \doteq & E_{\pi_{N}} E_{\theta}
\left[
 \left\{\hat{\psi}(D_{n} ; \pi_{I}) - \psi(\theta)\right\}^{2}
\right].
\end{eqnarray}
Note that we quite deliberately keep both the `A' and the `M' in `AMSE,' to remind us that two averagings are in play: the mean (M) of squared error across $D_n$ given $\theta$,
distinct from the averaging (A) of this result across $\theta$.

Standard decision-theoretic terminology
would see (\ref{AMSE}) referred to as the Bayes risk of the Bayesian estimator formed under $\pi_{I}$, with respect to $\pi_{N}$.
Indeed,
standard arguments tell us that for fixed $\pi_{N}$,
(\ref{AMSE}) is minimized by taking $\pi_{I}=\pi_{N}$,
and that this is the minimum achievable across {\em all} estimators,
not just those arising as Bayesian estimators from some choice of  prior
\citep[see, for instance,][Sec.\ 4.4]{berger1985statistical}.

We focus on the large-sample limit,
letting $AMSE() = \lim_{n \rightarrow \infty} AMSE_{n}()$.
The large-sample limit of the posterior mean,
$\psi^{\star}$,
depends on the true value of $\phi$ and the chosen prior
$\pi_{I}$,
e.g., $\psi^{\star}$ is given by (\ref{limzero}) if
$\pi_{I}=\pi_{0}$,
or by (\ref{limmix})
if $\pi_{I}=\pi_{MIX}$.
More formally then:
\begin{eqnarray} \label{limAMSE}
AMSE(\pi_{N},\pi_{I})
& \doteq & E_{\pi_{N}}
\left[
 \left\{\psi^{\star}(\phi ; \pi_{I})- \psi(\theta)\right\}^{2}
\right].
\end{eqnarray}
For interpretability we will report the square root
of $AMSE$, i.e., $RAMSE = AMSE^{1/2}$.

The ``two prior'' framework above (Nature and investigator)
deserves elaboration.
The frequentist MSE of the posterior mean is being averaged,
when (Nature's) weighting over $\Theta_0$ in forming this average may differ from the
(investigator's) weighting over $\Theta_0$ in determining the posterior given data.
While not seen commonly,
this approach has precedents.
For instance,
\citet{wang2002simulation} use the idea in a scheme to determine an appropriate sample size  (using ``simulation'' and ``fitting'' in place of ``Nature'' and ``investigator'').
Heuristically,
we can think of
the averaging with respect to $\pi_{N}$ as averaging across all the scenarios Nature {\em might} present to us.
Or, in a game-theory spirit,
we can think of Nature as choosing the true (but unrevealed) state of the world by drawing from $\pi_{N}$.
A different slant is espoused in \citet{gugr2009statsci},
whereby a quantity like
(\ref{AMSE}) reflects the long-run performance of a research lab tasked with investigating a sequence of {\em different} scientific relationships,
with $\pi_{N}$ describing variation across these relationships.

The two-prior approach can naturally quantify the pros and cons of asserting that perhaps an identified sub-model holds.
Particularly,
we can compute $AMSE(\pi_{N},\pi_{I})$ for the four combinations arising from
$\pi_{I} \in \{\pi_{0},\pi_{MIX}\}$
and
$\pi_{N} \in \{\pi_{0},\pi_{MIX}\}$.
Necessarily,
$AMSE(\pi_{MIX}, \pi_{MIX})$
is lower than (or possibly equal to)
$AMSE(\pi_{MIX}, \pi_{0})$.
The extent to which it is lower reflects the benefit of making a maybe assertion in an appropriate context,
since by setting $\pi_{N}=\pi_{MIX}$ we study average-case performance across a mix of scenarios,
some with,
and some without,
one of the identifying restrictions being true.

Conversely,
$AMSE(\pi_{0}, \pi_{MIX})$
is guaranteed to be higher than (or possibly equal to)
$AMSE(\pi_{0}, \pi_{0})$,
and the extent to which it is higher reflects the risk of  inappropriately making a maybe assertion.
That is, across scenarios where none of the identifying restrictions hold, making the maybe assertion increases average-case MSE.
Asserting maybe in a context where
all the identified sub-models are actually {\em a priori} implausible can be regarded as a form of cheating.
Thus the extent to which
$AMSE(\pi_{0}, \pi_{MIX})$
exceeds
$AMSE(\pi_{0}, \pi_{0})$
quantifies the reduction in estimation performance resulting from such cheating.

Note that the $AMSE$ comparisons just outlined could be applied in a fully identified setting
with $\psi$ identified under $M_{0}$ (and consequently also identified under each sub-model $M_{j}$, $j=1, \ldots J$).
For a finite sample size $n$, the four $AMSE_n$ values
determined by (\ref{AMSE}) would retain the interpretations given above,
quantifying both the value and the risk of speculating that the sub-models are {\em a priori} plausible.
However, (\ref{limAMSE}) would no longer be relevant.
All four $AMSE_n$ values would tend to zero as $n$ tends to infinity,
since both
$\hat{\psi}(D_n ; \pi_{0})$
and
$\hat{\psi}(D_n ; \pi_{MIX})$
would consistently estimate $\psi$,
under any values of $\theta \in \Theta_0$.
Conversely,
in the situation we study,
with $\psi$ not identified under $M_{0}$,
consistent estimation does not arise.
Hence the positive large-sample limits of the $AMSE$ governed by (\ref{limAMSE})
are fundamental descriptors of the situation.

\section{Example 1: Prevalence Estimation with Missing Data}

We return to the motivating example of Section 1.2.
Here $M_{0}$ is the NIM specification which does not fully identify the target, whereas $M_{1}$ is the MAR specification which does.
The prior specifications $\pi_{0}()$ and $\pi_{1}()$ are those given in Section 1.2.
The posterior mean of the disease prevalence $\psi$
has large-sample limit $\psi^{\star}_{0}=p^{\dagger}_{11} + p^{\dagger}_{0+}/2$
when the investigator allows NIM ($M_0$),
but $\psi^{\star}_{1}=p^{\dagger}_{11}/p^{\dagger}_{1+}$ when the investigator presumes MAR ($M_1$).

The respective prior
marginal densities of $\phi=(p_{0+},p_{10},p_{11})$ are readily obtained.
Expressed as densities for $(p_{0+},p_{11})$ over the lower triangle
of the unit square (rather than densities over the probability simplex), we have $\pi_{0}(p_{0+},p_{11})=6p_{0+}$ for the NIM specification,
and $\pi_{1}(p_{0+},p_{11})=(1-p_{0+})^{-1}$.
Consequently,
\begin{eqnarray}   \label{ex1wstar}
\frac{w^{\star}_{1}/w^{\star}_{0}}{w_{1}/w_{0}} &=&  \frac{1}{6p^{\dagger}_{0+}(1-p^{\dagger}_{0+})}.
\end{eqnarray}
This limiting Bayes factor could also be deduced by direct
inspection of
(\ref{bf}).
Note that (\ref{ex1wstar}) places us in Scenario 3 of the taxonomy given earlier in Section 2.1.

Based on the prior specifications and the limiting posterior forms,
and taking equal prior model weights $w=(0.5, 0.5)$,
we have the following analytic expressions:
\begin{eqnarray*}
AMSE(\pi_{0},\pi_{0}) &=&  E_{\pi_{0}}
\left\{  \frac{(p_{01}-p_{00})^{2}}{4} \right\} \\
&=& \frac{1}{40},
\end{eqnarray*}
and
\begin{eqnarray*}
AMSE(\pi_{MIX},\pi_{0}) &=&  (1/2) E_{\pi_{0}}
\left\{  \frac{(p_{01}-p_{00})^{2}}{4} \right\}
+ (1/2) E_{\pi_{1}}
\left\{  \frac{(p_{01}-p_{00})^{2}}{4} \right\}  \\
&=& \frac{1}{2}\left(\frac{1}{40}+\frac{1}{36}\right).
\end{eqnarray*}

The other two quantities needed take the form
$AMSE(\pi_{0},\pi_{MIX}) = E_{\pi_{0}} \left\{ w(p)^{2} \right\}$
and
$AMSE(\pi_{MIX},\pi_{MIX}) = (1/2)E_{\pi_{0}} \left\{ w(p)^{2} \right\}
+(1/2)E_{\pi_{1}} \left\{ w(p)^{2} \right\}$,
where
\begin{eqnarray*}
w(p) &=&
\frac{6p_{0+}p_{1+}}{1+6p_{0+}p_{1+}}
\left(p_{11}+\frac{p_{0+}}{2}\right) +
\frac{1}{1+6p_{0+}p_{1+}}
\left(\frac{p_{11}}{p1+}\right) - p_{+1}.
\end{eqnarray*}
Given the nonlinearity of $w()$, the needed expectations do not have analytic forms.
They are easily evaluated numerically,
however,
say via Monte Carlo draws from $\pi_{0}()$ and $\pi_{1}()$.

Table 1 gives the four $AMSE$ values.
Positively,
appropriate use of the maybe assertion
leads to a 16\% reduction in RAMSE, i.e.,
we see this reduction when Nature's prior is the mixture,
so we are averaging performance across some scenarios where the assertion holds and others when it does not.
Conversely,
inappropriate use of the maybe assertion leads to a
12\% increase in RAMSE.
This is seen when Nature's prior is $\pi_{0}$ alone,
so we are averaging performance exclusively across NIM scenarios.

\begin{table}
\caption{RAMSE values for different combinations of Nature's prior and the investigator's prior, in Example 1.\label{tab:tabone}
Values in the first column are exact.
Values in the second column are computed as
Monte Carlo averages
based on $10^5$ draws from each of $\pi_{0}()$ and $\pi_{1}()$.
The impact of the investigator using $\pi_{MIX}$ rather than $\pi_{0}$
is a 11.8\% increase in RAMSE
when this use is not warranted, but a 16.0\% decrease in RAMSE when it is warranted. The Monte Carlo standard errors for these two percentage changes are
0.3 percentage points and 0.2 percentage points, respectively.}

\begin{center}
\begin{tabular}{rrrr}
         &             & \multicolumn{2}{c}{Investigator} \\
         &             & $\pi_{0}$ & $\pi_{MIX}$ \\
 Nature  & $\pi_{0}$   & 0.158    & 0.177 \\
         & $\pi_{MIX}$ & 0.162    & 0.136
\end{tabular}
\end{center}
\end{table}

\section{Example 2: Estimating an Average Risk Difference from Stratified Data}

Consider three binary variables,
$(C,X,Y)$,
with $X$ an exposure variable, $Y$ an outcome variable, and $C$ a potential confounder.
The target parameter is presumed to be the average risk difference,
$\psi = E\{E(Y|X=1,C)-E(Y|X=0,C)\}$,
which can be motivated from a causal inference perspective.
Say the available data,
however,
are sampled conditional on $C$.
That is, pre-specified numbers of realizations, $n_{0}$ and $n_1$, are drawn from $(Y,X|C=0)$,
and $(Y,X|C=1)$ respectively.

Let $\phi=(\phi_{0},\phi_{1})$,
where $\phi_{c}$ comprises  the $(X,Y|C=c)$ cell probabilities, for $c=0,1$.
(So each $\phi_{c}$ is a probability vector, with four elements constrained to sum to one.)
Also,
let  $\lambda = Pr(C=1)$.
Then $(\phi, \lambda)$ is a transparent parameterization,
with the likelihood $f(D | \phi)$ factoring into $\phi_0$ and $\phi_1$ terms based on two independent multinomial samples.
The target parameter can be expressed as $g(\phi,\lambda) = (1-\lambda) v(\phi_{0}) + \lambda v(\phi_{1})$,
where $v()$ returns the risk difference from a single set of cell probabilities.
For the partially identified model $M_{0}$ we specify a prior with independencies of the form $\pi_{0}(\phi,\lambda) = \pi_{0}(\phi_{0})\pi_{0}(\phi_{1})\pi_{0}(\lambda)$.
For future reference,
let $\bar{\lambda}$ and $\sigma^{2}_{\lambda}$
be the mean and variance of $\lambda$ under the specified prior.
%
%
%
%

The first identifying restriction considered is that the prevalence of $C$ is exactly known from external sources,
i.e.,
$M_{1}$ is taken to be the sub-model of $M_{0}$ defined by $\lambda=\tilde{\lambda}$, where $\tilde{\lambda}$ is user-specified.
%
%
We express this model via the prior specification $\pi_{1}(\phi,\lambda) = \pi_{1}(\phi) \delta_{\tilde{\lambda}}(\lambda)$,
where $\delta_{x}()$ is the Dirac delta function, i.e., the
`density' of a point-mass at $x$.
We take $\pi_{1}(\phi)  = \pi_{0}(\phi)$ as before,
but now $\lambda=\tilde{\lambda}$ is taken as known.

The second identifying restriction considered is that of no interaction on the risk difference scale.
Hence we obtain $M_{2}$ from $M_{0}$ by constraining $\phi$  according to $v(\phi_{0})=v(\phi_{1})$.
An appropriate prior specification for the marginal $\pi_{2}(\phi)$ would be the distribution induced from $\pi_{0}(\phi)$ by
conditioning on $v(\phi_{0})=v(\phi_{1})$.
For $M_{2}$ we can leave $\pi_{2}(\lambda|\phi)$ unspecified,
since it will play no role whatsoever in the posterior distribution of $\psi$ or in the posterior weight of $M_{2}$ relative to the other two models.

From the forms of $\pi_{j}(\phi)$, for $j=0,1,2$, we have a very simple  characterization of the
limiting posterior model weights:
\begin{eqnarray} \label{ex1limwht}
(w^{\star}_{0}, w^{\star}_{1}, w^{\star}_{2}) &=&
\left\{
\begin{array}{ll}
(0,0,1) & \mbox{if $v(\phi_{0}^{\dagger}) = v(\phi_{1}^{\dagger})$;} \\
(w_{0}+w_{1})^{-1}(w_{0}, w_{1}, 0) & \mbox{otherwise.}
\end{array}
\right.
\end{eqnarray}
Specifically,
$M_{1}$ is completely untestable compared to $M_{0}$,
in the sense that $f_{0}(D)=f_{1}(D)$ for every dataset.
On the other hand $M_{2}$ is completely testable compared to $M_{0}$,
as $v(\phi_{0})=v(\phi_{1})$ is either proven or disproven in the limit of infinite sample size.
Referring back to the scenarios given in Section 2.1,
$M_1$ falls in the first scenario, while $M_{2}$ falls in the second.

To describe within-model inference,
under $M_{0}$ the posterior distribution of $(\psi|\phi)$ is a location-scale shift of the prior distribution of $\lambda$ [with location $v(\phi_{0})$ and scale $v(\phi_{1})-v(\phi_{0})$].
The limiting posterior mean of $\psi$ is then
$v(\phi^{\dagger}_{0}) + \{v(\phi^{\dagger}_{1})-v(\phi^{\dagger}_{0})\}\bar{\lambda}$.
Whereas,
relatedly,
under $M_{1}$ the limiting posterior mean is
$v(\phi^{\dagger}_{0}) + \{v(\phi^{\dagger}_{1})-v(\phi^{\dagger}_{0})\}\tilde{\lambda}$
(matching the target $\psi^{\dagger}$ if $M_{1}$ holds).
If $M_{2}$ holds,
then asymptotically the $M_{2}$ posterior will concentrate at the correct value of
$v(\phi_{0}^{\dagger})=v(\phi_{1}^{\dagger})$.
The limiting posterior mean under $M_{2}$ when $M_{2}$ does not hold is governed by standard wrong-model asymptotic theory
(e.g., see \citet{white1982maximum}).
We will not need to determine this limit,
however,
as (\ref{ex1limwht}) discredits $M_2$ when it is wrong.

Armed with (\ref{ex1limwht}) and the limiting posterior means of $\psi$ under each model,
we proceed to determine $AMSE$ for the four combinations of Nature's prior and the investigator's prior based on $\pi_{0}()$ or $\pi_{MIX}()$.
Letting $k=\mbox{Var}\{v(\phi_{1})-v(\phi_{0})\} = 2 \mbox{Var}\{ v(\phi_{c}\}$ under
$\pi_{0}()$,
direct calculation gives:
\begin{eqnarray*}
AMSE(\pi_{0} ; \pi_{0}) &=& k \sigma^{2}_{\lambda}, \\
AMSE(\pi_{0} ; \pi_{MIX}) &=& k \left[ \sigma^{2}_{\lambda} + \{w_{2}/(w_{1}+w_{2})\}^{2} \left(\tilde{\lambda}-\bar{\lambda}\right)^{2} \right], \\
AMSE(\pi_{MIX} ; \pi_{0}) &=&
k \left\{
w_{0}\sigma^{2}_{\lambda} + w_{1}\left(\tilde{\lambda}-\bar{\lambda}\right)^{2}
\right\}, \\
AMSE(\pi_{MIX} ; \pi_{MIX}) &=&
k \left\{
w_{0}\sigma^{2}_{\lambda} + \frac{w_{0}w_{1}}{w_{0}+w_{1}}\left(\tilde{\lambda}-\bar{\lambda}\right)^{2}
\right\}.
\end{eqnarray*}
These expressions are sufficiently simple that one can ``read off'' the
extents to which
$AMSE(\pi_{0}; \pi_{MIX})$ exceeds $AMSE(\pi_{0}; \pi_{0})$
and
$AMSE(\pi_{MIX}; \pi_{MIX})$
is reduced compared to
$AMSE(\pi_{MIX}; \pi_{0})$.
Note that both these gaps collapse to zero if $\bar{\lambda}$ (the prior mean
of $\lambda$ under model $M_{0}$) equals $\tilde{\lambda}$ (the presumed value of $\lambda$ under Model $M_{1}$).

To give a concrete example,
say external information leads to the specifying $\tilde{\lambda}=0.15$ for the
identifying restriction under $M_{1}$.
And the same source suggests the prior $\lambda ~ \sim \mbox{Beta}(4,18)$ under Model $M_{0}$, yielding the prior mode at $0.15$, with prior mean of
$\bar{\lambda}=4/22$
and prior variance of
$\sigma^{2}_{\lambda} = (4/22)(18/22)(1/23) \approx 0.00647$.
Using a uniform prior on cell probabilities,
i.e., $\phi_{c} \sim \mbox{Dirichlet}(1,1,1,1)$,
gives $k=1/3$.
Presuming equal prior weights on the models,
$(w_{0},w_{1},w_{2})=(1/3, 1/3, 1/3)$,
the resulting RAMSE values are given in Table 2.
These reveal a low-stakes setting.
Imposing the maybe restrictions when not warranted only carries a 1.9\% increase in RAMSE (first row).
While imposing them when warranted only carries a 3.5\% reduction in RAMSE (second row).

\begin{table}
\caption{RAMSE values for different combinations of Nature's prior and the investigator's prior in Example 2.
The impact of the investigator using $\pi_{MIX}$ rather than $\pi_{0}$ is a 1.9\% increase when this use is not warranted, but a 3.5\% decrease when it is warranted.}
\begin{center}
\begin{tabular}{rrrr}
         &             & \multicolumn{2}{c}{Investigator} \\
         &             & $\pi_{0}$ & $\pi_{MIX}$ \\
 Nature  & $\pi_{0}$   & 0.0464    & 0.0473 \\
         & $\pi_{MIX}$ & 0.0288    & 0.0278
\end{tabular}
\end{center}
\end{table}

\section{Example 3: Estimating a Risk Difference with Possible Outcome Misclassification}

In a similar spirit to Example 2,
say we are interested in the risk difference
$Pr(Y=1|X=1)-Pr(Y=1|X=0)$, for binary variables $X$ and $Y$.
However,
the available data are $(X,Y^{*})$ realizations,
where $Y^{*}$ may be an imperfect surrogate for $Y$, while $Y$ itself is latent.
Specifically,
say $Y^{*}$ and $X$ are conditionally independent given $Y$
(the so-called ``nondifferential'' assumption),
with $\lambda_{y}=Pr(Y^{*}=Y|Y=y)$,
i.e, $\lambda_{0}$ and $\lambda_{1}$ are respectively the specificity and sensitivity of the surrogate.
Letting $\omega_{x} = Pr(Y=1|X=x)$,
the target of inference is
$\psi = \omega_{1}-\omega_{0}$.
While $(\omega,\lambda)$ is an interpretable parameterization,
a transparent parameterization is $(\phi,\lambda)$,
where $\phi=(\phi_{0},\phi_{1})$, with
$\phi_{x}= Pr(Y^{*}=1|X=x) = (1-\omega_{x})(1-\lambda_{0}) + \omega_{x} \lambda_{1}$.
With respect to this parameterization,
the target is
$\psi= (\phi_{1}-\phi_{0})/(\lambda_{0}+\lambda_{1}-1)$.

%
%

As a prior specification for $M_0$,
we let $\omega$ follow a uniform distribution over $(0,1)^{2}$,
and independently let
$\lambda$ follow a uniform distribution over
$(a_{0},1) \times (a_{1},1)$.
Thus hyperparameters $a_{0}$ and $a_{1}$ are worst-case assertions about the
magnitude of outcome misclassification.
(We presume $a_{0}>0.5$ and $a_{1}>0.5$ for technical reasons,
which rules out placing any prior mass on a negative dependence between $Y$ and $Y^{*}$.)
The prior independence between $\omega$ and $\lambda$ is intuitive;
prior assertions about $(Y|X)$ prevalences are not tied to prior assertions about the quality of
$Y^{*}$ as a surrogate for $Y$.
(And prior independence of $\phi$ and $\lambda$ would not be possible,
since the support of $\lambda$ depends on $\phi$.)

Upon moving to the transparent parameterization,
following the related model formulation in \citet{gustafson2001biom},
the $M_{0}$ prior density transforms to
\begin{eqnarray} \label{ex2jointpri}
\pi_{0}(\phi,\lambda) &=&
\frac{I_{A}(\phi,\lambda)}{(1-a_{0})(1-a_{1})(\lambda_{0}+\lambda_{1}-1)^{2}}.
\end{eqnarray}
Here $A$ is the intersection of three sets,
imposing the restrictions that:
(i), $\phi \in (0,1)^{2}$;
(ii), $\lambda \in (a_{0},1) \times (a_{1},1)$;
and
(iii), $\phi$ and $\lambda$ are compatible with each other in that
$(1-\lambda_{0}) < \phi_{x} < \lambda_{1}$, for $x=0,1$.

The identifying restriction considered is the assertion of perfect outcome classification,
i.e., $M_{1}$ is the sub-model of $M_{0}$ defined by $\lambda=(1,1)^{\prime}$.
Under $M_{1}$ we take the prior on $\phi \sim \mbox{Unif}\{(0,1)^2\}$ {\em a priori}
This is consistent with $M_{0}$ in that it matches the $(\phi| \lambda=(1,1)^{\prime})$ prior;
see Appendix B for further discussion of prior specifications which match in this sense.

For a datastream arising under $\phi=\phi^{\dagger}$,
we have the limiting Bayes factor
$\pi_{1}(\phi^{\dagger})/\pi_{0}(\phi^{\dagger})$ as per (\ref{genlimbf}).
Clearly,
$\pi_{1}(\phi)=1$,
while marginalizing (\ref{ex2jointpri}) gives:
\begin{eqnarray} \label{tabtop}
\pi_{0}(\phi) &=&
\frac{1}{(1-a_{0})(1-a_{1})}
\left[
\log b_{0}(\phi) + \log b_{1}(\phi) - \log\{ b_{0}(\phi)+b_{1}(\phi)-1\}
\right],
\end{eqnarray}
where
$b_{0}(\phi)= \max\{ a_{0}, 1-\phi_{0}, 1-\phi_{1}\}$
and
$b_{1}(\phi)= \max\{ a_{1}, \phi_{0}, \phi_{1}\}$.
The bivariate density (\ref{tabtop}) over the unit square has a ``tabletop'' shape.
It is constant over the square
defined by
$\min\{\phi_{0},\phi_{1}\} > 1-a_{0}$
and
$\max\{\phi_{0},\phi_{1}\} < a_{1}$.
Whereas the density falls off away from this square,
with $\pi_{0}(\phi) \rightarrow 0$ as
$\min\{\phi_{0},\phi_{1}\} \rightarrow 0$ or
$\max\{\phi_{0},\phi_{1}\} \rightarrow 1$.

The shape of (\ref{tabtop}) has clear implications for what the data can say about the identifying restriction.
Say we are working with even prior models odds, $w=(0.5, 0.5)^{\prime}$.
Then,
for a datastream generated under a value of $\phi^{\dagger}$ within the tabletop mentioned above,
\begin{eqnarray*}
w_{1}^{\star}/w_{0}^{\star} &=&
\{(1-a_{0})(1-a_{1})\} /
[ \log \{a_{0}a_{1}/(a_{0}+a_{1}-1)\} ].
\end{eqnarray*}
This provides a bound on the extent of criticism the data can provide {\em against} the identified sub-model.
On the other hand,
there are values of $\phi^{\dagger}$ outside the tabletop
for which $w_{1}^{\star}/w_{0}^{\star}$ is arbitrarily large.
Thus circumstances exist under which the data can provide strong support {\em for} the identified sub-model.
(Though not as strong as if $M_{0}$ were identified,
in which case either $w^{\star}=(0,1)^{\prime}$ or $w^{\star}=(1,0)^{\prime}$.)
Referring back to taxonomy in Section 2.1,
the restriction $M_1$ falls into the third scenario.

\begin{figure}
\begin{center}
\includegraphics[width=5in]{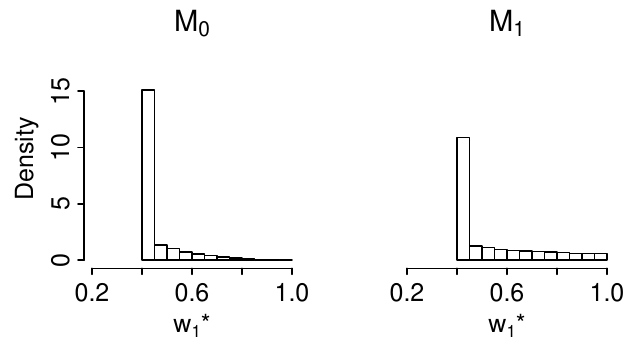}
\end{center}
\caption{Distribution of $w_{1}^{\star}$,
the limiting posterior weight on $M_{1}$,
in Example 3.
The panels correspond to parameter ensembles for which
$M_{0}$ (left) and $M_{1}$ (right) are true.
Hyperparameters are $a=(0.85, 0.85)^{\prime}$ and $w=(0.5,0.5)^{\prime}$.
}
\label{figfigEx3}
\end{figure}

To better convey the ability of data to discriminate between $M_{0}$ and $M_{1}$,
we
examine the $w^{\star}$ arising
for two ensembles of $\phi^{\dagger}$ values,
generated by sampling from $\pi_{0}()$ and $\pi_{1}()$, respectively.
The $M_0$ hyperparameter setting is
$(a_{0},a_{1})=(0.85, 0.85)$ for $\pi_{0}()$,
and equal {\em a priori} model weights, $(w_{0},w_{1})=(0.5, 0.5)$, are prescribed.
The two ensembles of the limiting posterior model weight $w^{\star}_{1}$
appear in Figure \ref{figfigEx3}.
As might be expected,
there is relatively more mass at the lower boundary for $w^{\star}_{1}$
when $M_{1}$ is false.
(For the present hyperparameter values this lower boundary is $0.416$,
representing a very limited scope for criticism of $M_{1}$.)
As per the discussion above,
the right tail of the $w^{\star}_{1}$ values approaches one when $M_{1}$ is true, i.e., there are occasional circumstances where
the data can strongly support the identifying restriction.

In terms of limiting posterior means for the target,
when the investigator presumes $M_{1}$,
the limit is simply
$\psi^{\star}_{1}=\phi^{\dagger}_{1} - \phi^{\dagger}_{0}$.
For $M_{0}$, numerical integration is required to
obtain $\psi^{\star}_{0}$ for a given $\phi^{\dagger}$.
Specifically, $\psi^{\star}_{0}= \int g(\phi^{\dagger},\lambda)
\pi_{0}(\lambda|\phi^{\dagger}) d\lambda$,
where the prior conditional density for $(\lambda | \phi)$ is derived from the joint density (\ref{ex2jointpri}).

The RAMSE values, given in Table 3, reveal higher stakes than manifested in Example 2.
Imposing the maybe restriction when not warranted incurs a penalty of a
45\% increase in RAMSE (first row).
Whereas imposing the maybe restriction when warranted
produces a 25\% reduction in RAMSE (second row).

\begin{table}
\caption{RAMSE values for different combinations of Nature's prior and the investigator's prior in Example 3.
Each value is computed as a Monte Carlo average,
using $20000$ realizations from each of $\pi_{0}()$ and $\pi_{1}()$.
The impact of the investigator using $\pi_{MIX}$ rather than $\pi_{0}$ is a 44.8\% increase in RAMSE
when this use is not warranted, but a 24.9\% decrease in RAMSE when it is warranted.
Monte Carlo standard errors for these two percentage changes are
$0.9$ percentage points and $0.3$ percentage points, respectively.
}
\begin{center}
\begin{tabular}{rrrr}
         &             & \multicolumn{2}{c}{Investigator} \\
         &             & $\pi_{0}$ & $\pi_{MIX}$ \\
 Nature  & $\pi_{0}$   & 0.0270    & 0.0391 \\
         & $\pi_{MIX}$ & 0.0462    & 0.0347  \\
\end{tabular}
\end{center}
\end{table}

\section{Example 4: Estimating an Average Risk Difference with Possible Exposure Misclassification}

Our final, and most complex, example blends elements of Examples 2 and 3.
As per Example 2,
consider the joint distribution of $(C,X,Y)$,
where $C$ is a binary confounding variable,
$X$ is a binary exposure variable,
and $Y$ is a binary outcome variable.
The target of inference
is again presumed to be the average risk difference, $\psi=E\{Pr(Y=1|X=1,C) - Pr(Y=1|X=0,C)\}$.
However,
the exposure $X$ may be subject to misclassification.
Hence the observable variables are $(C,X^{*},Y)$,
where the surrogate exposure $X^{*}$ may differ from $X$.
We confine attention to a simple form of exposure misclassification.
We take it as known that the misclassification is nondifferential,
in the sense that $X^{*}$ is conditionally independent of $(C,Y)$ given $X$.
Moreover, we presume $Pr(X^{*}=0|X=0)=1$,
i.e., the classification scheme has perfect {\em specificity}.
However,
the {\em sensitivity},
$\lambda = Pr(X^{*}=1|X=1)$,
may be less than one.
Some recent work on ``unidirectional'' misclassification of this form
includes
\citet{Xia2016, xia2018bayesian}.

\subsection{Convenience Prior and Posterior}

To organize prior specification and posterior computation across multiple models,
it is helpful to specify a {\em convenience} prior distribution which in turn produces a
convenience posterior distribution with a simple form.
Appropriate model-specific prior specifications and posterior calculations can be expressed as
tweaks to
the convenience analysis.

Let ${\cal P}_{d}$ be the space of probability vectors over
$d$ mutually distinct and exhaustive outcomes,
and let $\phi \in {\cal P}_{8}$ be the cell probabilities
describing the distribution of $(C,X^{*},Y)$.
We specify the convenience prior density on $\theta=(\phi,\lambda)$
to be $\pi_{*}(\phi,\lambda)=\pi_{*}(\phi) \pi_{*}(\lambda)$,
where $\pi_{*}(\phi)$ is the $\mbox{Dirichlet}(1, \ldots, 1)$ density,
while $\pi_{*}(\lambda)$ is the $\mbox{Uniform}(b,1)$ density,
where hyperparameter $b$ is an {\em a priori} specified lower bound on the sensitivity of the exposure classification.
For illustration,
we take $b=0.5$ throughout.

\subsection{Model $M_{0}$}
In the absence of any identifying restrictions,
a first thought is that the convenience prior might be employed as the actual prior.
This is not actually possible,
however,
since we do {\em not} have a Cartesian parameter space for $(\phi,\lambda)$.
For a given sensitivity $\lambda$,
let $s_{\lambda}()$ map from the $(C,X,Y)$ cell probabilities to the $(C,X^{*},Y)$ cell probabilities.
Hence $s_{\lambda}()$ maps from ${\cal P}_{8}$ to its image ${\cal S}_{\lambda} \subset {\cal P}_{8}$.
This map is invertible,
and for a given $\phi \in {\cal P}_{8}$ it is possible to numerically determine whether $\phi \in {\cal S}_{\lambda}$,
and,
if so,
compute $s^{-1}_{\lambda}(\phi)$.
Thus an obvious adaptation of the convenience prior is to take the Model $M_{0}$
prior as:
\begin{eqnarray} \label{pri00}
\pi_{0}(\phi,\lambda) & = &
\frac{\pi_{*}(\phi,\lambda) I\{\phi \in S_{\lambda}\}}
     { Pr_{*} \{\phi \in S_{\lambda}\} }.
\end{eqnarray}
Thus we simply truncate the convenience prior to only those $\phi$ and $\lambda$ pairs which are compatible with each other.

It transpires that
$\phi \in {\cal S}_{\lambda}$ if and only if $\lambda$ exceeds a threshold $t(\phi)$,
i.e.,
the cell probabilities for the observables $(C,X^{*},Y)$ imply a lower bound on the sensitivity of $X^{*}$ as a surrogate for $X$.
(Note that this bound may or may not exceed the
{\em a priori} lower bound $b$.)
Combined with the specification of
$\pi_{*}(\lambda)$ as $\mbox{Unif}(b,1)$,
(\ref{pri00}) can be simplified to:
\begin{eqnarray*}
\pi_{0}(\phi,\lambda) & = &
\frac{ \pi_{*}(\phi,\lambda) I[\lambda > \max\{b,t(\phi)\}]}
     { (1-b)^{-1}E_{*}[1-\max\{b,t(\phi)\}] }.
\end{eqnarray*}
Importantly,
this marginalizes to
\begin{eqnarray} \label{Ex3Marg0}
\pi_{0}(\phi) & = &
\frac{ \pi_{*}(\phi) [1- \max\{b,t(\phi)\}] }
     { E_{*}[1-\max\{b,t(\phi)\}] }.
\end{eqnarray}

In terms of the parameter of interest,
say that $\tilde{g}()$ maps from the $(C,X,Y)$ cell probabilities to the target parameter,
i.e.,
$\tilde{g}()$ returns $E\{E(Y|X=1,C)-E(Y|X=0,C)\}$.
In the parameterization at hand then,
the target parameter is $g(\phi,\lambda) = \tilde{g}( s_{\lambda}^{-1}(\phi))$.

\subsection{Model $M_{1}$}

In a similar spirit to Example 3,
the first identifying restriction considered is simply that the surrogate $X^{*}$ is in fact perfect.
So model $M_{1}$ is the sub-model of $M_{0}$ corresponding to $\lambda=1$.
An obvious prior specification is
\begin{eqnarray*}
\pi_{1}(\phi,\lambda) & = & \pi_{*}(\phi) \delta_{1}(\lambda).
\end{eqnarray*}
This simply marginalizes to
\begin{eqnarray} \label{Ex3Marg1}
\pi_{1}(\phi) & = & \pi_{*}(\phi).
\end{eqnarray}

Under this restriction $\phi$ is identically the $(C,X,Y)$ cell probabilities.
Hence the target parameter is simply expressed as $g(\phi)=\tilde{g}(\phi)$.

\subsection{Model $M_{2}$}

In the spirit of Example 2,
Model $M_{2}$ renders the target identifiable via
asserting an absence of interaction on the risk difference scale,
so that $E(Y|X=1,C=c)-E(Y|X=0,C=c)$ does not depend on $c$.
Letting $\tilde{\phi}$ parameterize the so restricted $(C,X,Y)$ cell probabilities
(so that $\tilde{\phi}$ has only six degrees of freedom),
the resultant $(C,X^{*},Y)$ cell probabilities can be expressed as $h(\tilde{\phi},\lambda)$.
Here the map $h()$ is invertible.
However,
its image,
which we denote as ${\cal H}$,
is a strict subset of ${\cal P}_{8}$.
While it does not seem possible to express $h^{-1}()$ in closed form,
for a given $\phi \in {\cal P}_{8}$ one can numerically
determine whether $\phi \in {\cal H}$,
and,
if so,
compute $h^{-1}(\phi)$.


A sensible prior construction for $M_{2}$ thus takes the form:
\begin{eqnarray*}
\pi_{2}(\phi,\lambda) &=&
\frac{\pi_{*}(\phi) I_{{\cal H}}(\phi)}{ Pr_{*}(\phi \in {\cal H})}
\delta_{m(\phi)}(\lambda),
\end{eqnarray*}
where $m(\phi)$ is the unique sensitivity value implied by $\phi$.
Clearly this prior marginalizes to
\begin{eqnarray} \label{Ex3Marg2}
\pi_{2}(\phi) &=&
\frac{\pi_{*}(\phi) I_{{\cal H}}(\phi)}{ Pr_{*}(\phi \in {\cal H})}.
\end{eqnarray}
%

%

\subsection{Limiting Inference across models}

The limiting posterior model weights are governed by
(\ref{Ex3Marg0}), (\ref{Ex3Marg1}), and (\ref{Ex3Marg2}),
giving
\begin{eqnarray} \label{Ex3Asymp1}
\frac{w_{0}^{\star}/w_{1}^{\star}}{w_{0}/w_{1}}
&=&
\frac{ 1-\max\left\{ b, t\left(\phi^{\dagger}\right) \right\} }
{ 1-E_{*}[\max\{ b, t(\phi) \}] },
\end{eqnarray}
and
\begin{eqnarray} \label{Ex3Asymp2}
\frac{w_{2}^{\star}/w_{1}^{\star}}{w_{2}/w_{1}}
&=&
\left \{
\begin{array}{ll}
\{Pr_{*}( \phi \in {\cal H} )\}^{-1} & \mbox{if $\phi^{\dagger} \in {\cal H}$}, \\
0 & \mbox{otherwise.}
\end{array}
\right.
\end{eqnarray}
Via (\ref{Ex3Asymp1}) the data can offer some mild discrimination between
$M_{0}$ and $M_{1}$,
in that a positive and finite limiting Bayes factor arises.
Moreover,
this limit is not one (except for a set of $\phi^{\dagger}$ values
with probability zero under all three model-specific priors).
(Of course this is indeed mild.
If $M_0$ were identified, the limiting Bayes factor would be zero or infinity.)
Via (\ref{Ex3Asymp2}),
when $M_{2}$ is false,
enough data may or may not discredit
it.
That is,
the true model (either $M_{0}$ or $M_{1}$) may give rise to $\phi^{\dagger} \notin {\cal H}$,
in which case $M_{2}$ is discredited asymptotically.
However,
any of the three models can yield $\phi^{\dagger} \in {\cal H}$,
in which case (\ref{Ex3Asymp2})
offers no discrimination.
Note that if $M_{2}$ is not discredited,
then it necessarily receives more support that $M_{1}$,
i.e., (\ref{Ex3Asymp2}) exceeds one.
Linking back to the Section 2.1 taxonomy,
$M_1$ and $M_2$ are instances of Scenarios 3 and 4, respectively.

\subsection{Performance}

To further convey the ability to discriminate between the three models,
we examine $w^{\star}$ values arising for three ensembles of $\phi^{\dagger}$ values,
where the ensembles are randomly drawn from $\pi_{0}(\phi)$,
$\pi_{1}(\phi)$,
and $\pi_{2}(\phi)$ respectively.
Equal prior weights $w=(1/3,1/3/1/3)$ are employed.
The results are plotted in Figure 3.
For the ensembles corresponding to $M_{0}$ and $M_{1}$,
$w^{\star}_{2}=0$ for large majorities of points.
So $M_{2}$ is often,
but not always,
fully discredited when it is false.
On the other hand, when $M_{2}$ is not discredited (the minorities of scenarios arising under $M_{0}$ and $M_{1}$, but {\em all} the scenarios arising under $M_{2}$), it receives strong support.
The $w^{\star}_{2}$ values range from $0.72$ to $0.92$.

Focusing on the $w^{\star}_{2}=0$ cases generated under $M_{0}$ and
$M_{1}$,
we see a modest tendency for $w_{1}^{\star}=1-w^{\star}_{0}$ to be smaller  when $M_{0}$ is true and larger when $M_{1}$ is true.
There is also an asymmetry.
The strongest evidence
in favour of $M_{0}$ corresponds to a value of
$w^{\star}_{0}=1-w^{\star}_{1}$ well below one (approximately $w^{\star}_{0}=0.73$), and this extreme can be reached under either $M_{0}$ or $M_1$.
Conversely,
there are narrow circumstances under which
$w^{\star}_{1}=1-w^{\star}_{0}$ is very close to one.
As was the case in Example 3, there is more scope to support the identifying restriction $M_{1}$ than
there is to criticize it.

\begin{figure}
\begin{center}
\includegraphics[width=5in]{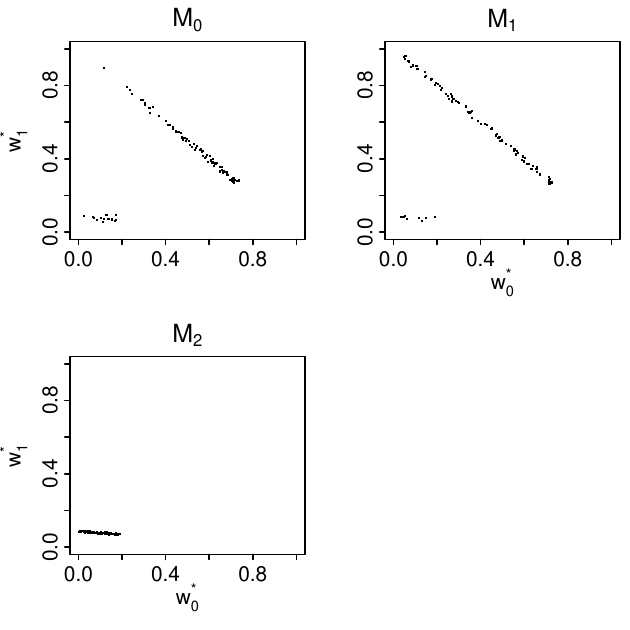}
\end{center}
\caption{Distribution of the limiting posterior weights
$(w^{\star}_{0},w^{\star}_{1},w^{\star}_{2})$, in Example 4,
under $M_{0}$ (upper-left), $M_{1}$ (upper-right), and $M_{2}$ (lower-left).
In each instance, an ensemble of
100 $\phi^{\dagger}$ points are simulated from $\pi_{j}()$.
Points in the upper panels are jittered with a small amount of random noise, in order to better see the distribution of those points with $w^{\star}_{0}+w^{\star}_{1}=1$, $w^{\star}_{2}=0$.}
\end{figure}

The RAMSE values are given in Table 4.
They show fairly high stakes attached to invoking the maybe assumptions.
RAMSE is increased by 31\% in terms of average performance across scenarios where neither restriction holds.
Whereas RAMSE is reduced by 40\% in terms of average performance
when the mix of scenarios indeed has one of the
restrictions holding in a portion of cases.

\begin{table}
\caption{RAMSE values for different combinations of Nature's prior and the investigator's prior in Example 4.
Each value is computed as a Monte Carlo average,
using 1600 draws from each of $\pi_{0}()$, $\pi_{1}()$, $\pi_{2}()$.
The impact of the investigator using $\pi_{MIX}$ rather than $\pi_{0}$
is a 31.3\% increase in RAMSE
when this use is not warranted, but a 39.5\% decrease in RAMSE when it is warranted. Monte Carlo standard errors for these two percentage changes are
$4.0$ percentage points and $2.1$ percentage points, respectively.}

\begin{center}
\begin{tabular}{rrrr}
         &             & \multicolumn{2}{c}{Investigator} \\
         &             & $\pi_{0}$ & $\pi_{MIX}$ \\
 Nature  & $\pi_{0}$   & 0.0452    & 0.0594 \\
         & $\pi_{MIX}$ & 0.0778    & 0.0470
\end{tabular}
\end{center}
\end{table}

\section{Discussion}

In current applied practice,
when faced with a lack of full identification unless an identifying but untestable assumption is imposed,
one generally imposes the restriction or not.
In the terminology of this paper,
and with only a single restriction contemplated,
this is choosing hyperparameter settings of either $w=(0,1)$ or $w=(1,0)$.
To our knowledge,
this has been the first
investigation of intermediate settings, such as $w=(0.5, 0.5)$,
with the interpretation of asserting that perhaps the restriction holds.

We certainly do not claim a free lunch arises from saying perhaps.
But we do think it a worthwhile analysis strategy to consider,
simply because many applied problems do arise where an identifying restriction
is both plausible,
but not incontrovertible,
in the scientific context at hand.
Moreover,
the finding that sometimes the limiting posterior model weights differ from
the prior model weights adds intrigue,
and additional promise,
to the use of perhaps assertions.

In general, to specify a prior distribution
is to implicitly choose the estimator with optimal average-case behavior, where the average is with respect to the joint distribution of parameters and data arising from the specified prior.
In situations with plausible identifying restrictions,
we have seen the stakes associated with prior assertions can be quite high.
In one direction,
$AMSE(\pi_{0},\pi_{MIX})$ can be substantially higher
than $AMSE(\pi_{0},\pi_{0})$.
Purely wishful thinking that perhaps an identifying restriction holds
can come with a steep penalty.
Equally,
however,
$AMSE(\pi_{MIX},\pi_{MIX})$ can be substantially lower
than $AMSE(\pi_{MIX},\pi_{0})$.
Realistic assessment that perhaps an identifying restriction holds comes with a reward.

An admitted limitation of the work is the presumption that $M_0$ is not violated.
Put plainly, we admit we might lack identification,
but in doing so we presume no model misspecification.
This issue is moot in Examples 1 and 2,
since $M_0$ does not place restrictions on the underlying $(R,Y)$ distribution in the former instance, or the underlying $(C,X,Y)$ distribution in the latter.
In Examples 3 and 4, however, $M_0$ involves a risk of misspecification,
particularly through assumptions about misclassification mechanisms and {\em a priori} bounds on misclassification parameters.

Finally,
returning to the posterior updating of model weights,
we have seen identifying restrictions that are empirically untestable,
in the sense that $w_{1}^{*}/w_{0}^{\star}$ is neither zero nor infinite,
i.e., even an infinite amount of data would neither definitively prove nor definitively disprove the identifying assumption.
In some such cases,
however,
the positive and finite value of $w_{1}^{*}/w_{0}^{\star}$
does vary with the underlying parameter values
$\theta^{\dagger}$ (via dependence on $\phi^{\dagger}$).
Commensurately,
in finite samples the posterior model weights can depend on the data.
Thus a nuanced situation of partial learning about the plausibility of a restriction can result.
Consequently,
the pros and cons of making perhaps suppositions {\em a priori} are not easily and generally intuited.
For a given problem,
however,
we have shown how to elucidate the structure of the inference.
Consequently,
the risks and rewards of giving some prior credence to one or more identifying restrictions can be quantified.

\section*{Supplementary Materials}

R code to produce all the empirical results in the paper are available
from \url{github.com/paulgstf/ParameterRestrictionsPerhaps}.

\section*{Appendix A: Sensitivity to Hyperparameters}

Here we elaborate on sensitivity to the prior specification,
in the context of the motivating example explored in Section 1.1 and Section 3.
Say we retain the $M_0$ within-model prior $p \sim \mbox{Dirichlet}(1,1,1,1)$.
Under the MAR assumption ($M_1$),
however,
we generalize from
$\gamma,\psi \stackrel{{\small \mbox{iid}}}{\sim} \mbox{Beta}(1,1)$
to
$\gamma,\psi \stackrel{{\small \mbox{iid}}}{\sim} \mbox{Beta}(k,1)$
for some choice of hyperparameter $k$.
(See Appendix B for a rationale for the choice $k=2$.)
%

%
%
%
%

%

In general,
regardless of identification issues,
Bayes factors are known to be sensitive to the choice of prior specifications within the two models being compared.
This is certainly the case in the present situation.
%
%
The ``conjugate integration'' that gave the Bayes factor (\ref{bf}) generalizes to
\begin{eqnarray} \label{bfgen}
b_{k} &=&
\frac{k^{2}}{6}
\frac{ (n+3)!}{ \Gamma(n+k+1) }
\frac{\Gamma(c_{11}+k)}{c_{11}!}
\frac{1}{(c_{0+}+1)(n-c_{0+} + 1)}.
\end{eqnarray}
For the dataset from Section 1.1,
the  Bayes factor is indeed very sensitive to increasing $k$ above the base value of $k=1$ (e.g., $b_{1.5}=2.17$, $b_{2}=1.00$, compared to  $b_{1}=3.73$).
In the other direction,
there is less sensitivity to decreasing $k$
(e.g., $b_{0.75}=4.14$, $b_{0.5}=3.62$).
%


Importantly,
alternative values of the hyperparameter $k$ also result in a Bayes factor which responds sensitively to the observed data,
a key point in the earlier developments.
The case $k=2$ is instructive,
as it yields a simple expression for (\ref{bfgen}),
namely
$b_{2} =  (2/3) \{(c_{11}+1)/(c_{1+}+1)\} / \{c_{0+}/n\}$,
with the latter two terms being, approximately,
the sample proportion of $Y=1$ amongst responders, and the sample proportion of non-responders.
So we see again a strong dependence on the data,
the lack of identification notwithstanding.
However, the nature of this dependence is rather different than
for $k=1$.

\section*{Appendix B: Prior specifications for nested models}

With Bayesian model averaging,
there is no overarching prescription for how respective within-model prior specifications should relate
when one model is nested within another.
Following our notation of $M_1$ being nested within $M_0$,
say $\pi_{0}()$ is already specified.
\citet{kass1995bayes}
point out that specifying $\pi_{1}()$ via either marginalizing or conditioning has intuitive appeal
(though there aren't strictures on other possibilities).
To be more explicit,
presume an initial parameterization of $M_0$ of the form
$\theta=(\theta_{a},\theta_b)$, with $M_1$ defined by $\theta_b=0$.
Then marginalizing leads to
$\pi_{1}(\theta_a) = \int \pi_{0}(\theta_a,\theta_b) d\theta_b$,
while conditioning leads to
$\pi_{1}(\theta_{a}) = \pi_{0}(\theta_a|\theta_b=0)$.
Many of our examples use the conditioning route.
Particularly, this is true in Example 1 (with $\theta_b=p_{11} - (p_{01}+p_{11})(p_{10}+p_{11})$),
Example 2 (with $\theta_b=\lambda-\tilde{\lambda}$ and $\theta_b= \nu(\phi_1)-\nu(\lambda_0)$ for the two nested models respectively),
and Example 3 (with $\theta_b=\lambda=(1,1)^{\prime}$).


\citet{kass1995bayes},
citing an unpublished communication chain with J.\ Dickey and L.J.\ Savage,
raise an interesting point about the conditioning route being
subject to the ``Borel paradox.''
This is indeed somewhat more than a theoretical curiosity.
For instance, consider Example 1.
Starting from $\pi_{0}()$
of the form
$p \sim \mbox{Dir}(1,1,1,1)$,
instantiating the MAR assumption `additively' by
conditioning on
$p_{11}-(p_{01}+p_{11})(p_{10}+p_{11})=0$ results in
$\gamma,\psi \stackrel{{\small \mbox{iid}}}{\sim} \mbox{Beta}(1,1)$,
as used in Sections 1.3 and 3.
However, instead instantiating MAR `multiplicatively' by conditioning on
$p_{11}(p_{01}+p_{11})^{-1}(p_{10}+p_{11})^{-1}-1=0$ leads to
$\gamma,\psi \stackrel{{\small \mbox{iid}}}{\sim} \mbox{Beta}(2,1)$,
as used for sensitivity analysis in Appendix A.

\bibliographystyle{Chicago}
\bibliography{../../../../../../ownCloud/REFS/mybib}

\end{document}